\documentstyle[12pt]{article}

\def\eqt#1{Eq.(\ref{#1})}
\def\eqts#1{Eqs.(\ref{#1})}
\def\eqtm#1{(\ref{#1})}
\def\hei{\vphantom{\Bigg|}}

\def\ket#1{\left|#1\right\rangle}
\def\braket#1#2{\left\langle#1\vphantom{#2}\right|
		\left.\vphantom{#1}#2\right\rangle}

\def\alt{\mathrel{\mathpalette\vereq<}}

\def\vereq#1#2{\lower3pt\vbox{\baselineskip1.5pt \lineskip1.5pt
               \ialign{$\mpth#1\hfill##\hfil$\crcr#2\crcr\sim\crcr}}}
\def\mpth{\mathsurround=0pt}
\def\erm{{\rm e}}

\def\epsilonslash{\epsilon\kern-.4em/}
\def\kslash{k\kern-.47em/}
\def\Lslash{L\kern-.45em/}
\def\pslash{p\kern-.435em/}
\def\partialslash{\partial\kern-.53em/}
\def\qslash{q\kern-.46em/}
\def\Rslash{R\kern-.6em/}
\def\sslash{s\kern-.44em/}
\def\vslash{v\kern-.47em/}
\def\GeV{\,{\rm G e\hskip-1pt V}}

\def\pvet{\vec{\hskip+1pt p}\hskip+1pt}

\newcounter{refs}
\begin{document}
\flushbottom
\pagestyle{empty}
\setcounter{page}{0}
\rightline{ DFTT 66/92 }
\rightline{ hep-ph/9211269 }
\rightline{ November 1992 }
\vskip1cm
\centerline{\LARGE\bf See-saw type mixing and
$ \nu_{\mu} \to \nu_{\tau} $ oscillations }
\vskip1cm
\centerline{\Large S.M. Bilenky }
\medskip
\centerline{\large\it Joint Institute of Nuclear Research }
\centerline{\large\it Dubna, Russia }
\centerline{\large\it and }
\centerline{\large\it
Istituto Nazionale di Fisica Nucleare, Sezione di Torino }
\centerline{\large\it
Via P. Giuria 1, I--10125 Torino, Italy }
\vskip0.5cm
\centerline{\Large C.~Giunti $^{\star}$ }
\medskip
\centerline{\large\it
Istituto Nazionale di Fisica Nucleare, Sezione di Torino }
\centerline{\large\it
Via P. Giuria 1, I--10125 Torino, Italy }
\vspace*{0.5in}
\centerline{\Large Abstract }
\bigskip

We consider
$ \nu_{\mu} \to \nu_{\tau} $
oscillations
under the assumption
that there is a see-saw type mixing
of the light neutrinos with
heavy Majorana particles.
It is shown
that the existing data,
including the recent LEP data,
do not exclude the possibility
that the additional terms
in the transition probability
due to this mixing
could be of the same order of magnitude as the
usual oscillating term.
Detail investigations
of $ \nu_{\mu} \to \nu_{\tau} $ transitions
in future CERN and Fermilab experiments
could allow to get informations
not only about the neutrino masses and mixing
but also about the mixing of neutrinos
with heavy Majorana particles.

\vfill
{\footnotesize
$^{\scriptstyle\star}$ Bitnet address: GIUNTI@TORINO.INFN.IT }
\newpage
\pagestyle{plain}
The problem of neutrino masses and mixing
is one of the main challenges of today's physics.
Many experiments devoted to this problem
have been performed recently.
Up to now
the only indications in favour of the existence
of a neutrino mass come
from the solar neutrino data [\ref{SolarData}]
and, possibly, from the atmospheric neutrino data [\ref{AtmData}].

At present a new generation of
solar neutrino experiments [\ref{NewSolarExp}],
experiments for search of neutrinoless double beta decay,
short and long baseline neutrino oscillation experiments
with reactor and accelerator neutrinos [\ref{Schneps92}]
and other experiments
are under development.
Two new
$ \nu_\mu \to \nu_\tau $
oscillation experiments
with a high sensitivity to small mixing angles
are under preparation at CERN [\ref{CHORUS},\ref{NOMAD}].
Analogous experiments
are planned at Fermilab [\ref{FNAL}].

Having in mind
these future experiments,
we consider here
$ \nu_\mu \to \nu_\tau $
oscillations
under the assumption that
the light neutrinos
mix with heavy Majorana particles.
This scenario corresponds to many models
beyond the standard model.

In accordance with the neutrino mixing hypothesis
[\ref{Bilenky78}],
the fields $ \nu_{\ell L} $
which appear
in the standard charged and neutral weak currents
\begin{equation}
\begin{array}{lll} \displaystyle
j^{W}_{\alpha}
& = & \displaystyle
2 \sum_{\ell = e,\mu,\tau}
\bar\nu_{\ell L} \, \gamma_{\alpha} \, \ell_{L}
+ \ldots
\\ \displaystyle
j^{Z}_{\alpha}
& = & \displaystyle
\sum_{\ell = e,\mu,\tau}
\bar\nu_{\ell L} \, \gamma_{\alpha} \, \nu_{\ell L}
+ \ldots
\end{array}
\end{equation}
are mixtures of the left-handed components
$ \nu_{aL} $
of the massive neutrino fields:
\begin{equation}
\nu_{\ell L} = \sum_{a} U_{\ell a} \nu_{aL}
\;,
\label{E02}
\end{equation}
where $U$ is a mixing matrix
which satisfy the unitarity relation
\begin{equation}
\sum_{a} U_{\ell a} U_{\ell' a}^* = \delta_{\ell\ell'}
\;.
\label{E03}
\end{equation}

    From the LEP data
it follows that the number of light neutrino flavours
is equal to 3 [\ref{LEP92}].
As for the number of massive neutrinos,
the LEP data do not allow to exclude different theoretical possibilities.
If massive neutrinos are Dirac particles (Dirac mass term),
their number is equal to the number of lepton flavours
[\ref{Bilenky78}].
In the case of massive Majorana neutrinos,
there are two different possibilities.
In the simplest case (Majorana mass term)
the number of neutrinos with definite mass
is equal to the number of lepton flavours.
In the most general case of neutrino mixing
(Dirac and Majorana mass term)
the number of massive Majorana particles
is more than 3.
This general type of mixing
has a great theoretical interest
and arises in many models beyond the standard model
[\ref{Bilenky78},\ref{Langacker91}].
The see-saw mechanism [\ref{GRSY}],
which seems to be the most plausible
mechanism for the generation of neutrino masses,
could be realized in this scheme.
In accordance with this mechanism,
the smallness of the masses of the light neutrinos
is due to their mixing with very heavy Majorana fermions
which have masses much larger than the masses of the charged
fermions (quarks or leptons).
There are many models in which the see-saw mechanism
for the generation of neutrino masses
take place [\ref{Langacker91}].
They differ mainly in the assumptions
about the heavy Majorana sector.

Here we consider neutrino oscillations
under the assumption
that the current fields
$ \nu_{\ell L} $
are mixtures of 3 fields of light Majorana neutrinos
with masses $ m_i $ ($i=1,2,3$)
and some fields of heavy Majorana particles
with masses $ m_a $ ($a=4,\ldots$)
at least larger than the mass of the $Z$ boson.
We do not assume any specific see-saw model
and consider the non-diagonal matrix elements
$ U_{\ell a} $ ($a\ge4$)
(which are small in most see-saw models)
as parameters.
It is clear that any information about these matrix elements
which can be obtained from experiment
is very important for understanding the nature of the neutrino mass.

Let us consider
neutrinos with momentum $\pvet$
which are produced together with charged leptons
$ \ell^{+} $ ($\ell=e,\mu,\tau$)
in some charged-current weak decay.
In accordance with \eqt{E02},
the neutrinos are described by the state
\begin{equation}
\ket{ \nu_{\ell} }
=
\sum_{i=1,2,3} U_{\ell i}^* \ket{ i }
\;,
\label{E04}
\end{equation}
where $\ket{i}$
are the states of the light neutrinos
with mass $ m_i \ll p $.
Let us stress that
the sum in \eqt{E04}
is only over the light mass eigenstates
(the heavy mass eigenstates cannot be produced).
After some time $t$ the state of the beam is given by
\begin{equation}
\ket{ \nu_{\ell} }_{t}
=
\sum_{i=1,2,3} U_{\ell i}^* \, \erm^{-iE_{i}t} \ket{ i }
\label{E04t}
\;,
\end{equation}
where
$ \displaystyle
E_{i} \simeq p + { m_i^2 \over 2 p } $.
Neutrinos are analyzed with weak interaction processes.
So the transition amplitude for
$ \nu_{\ell} \to \nu_{\ell'} $
is given by
\begin{equation}
A \left( \nu_{\ell} \to \nu_{\ell'} \right)
=
\braket{ \nu_{\ell'} }{ \nu_{\ell} }_{t}
=
\sum_{i=1,2,3} U_{\ell' i} \, \erm^{-iE_{i}t} \, U_{\ell i}^*
\label{E05}
\;.
\end{equation}
This expression has the same form as the usual
expression for the transition amplitude [\ref{Bilenky78}].
The essential difference is that
the sum in \eqt{E05} is only
over the indices which correspond to the light neutrinos.
In order to see more clearly this difference,
let us rewrite \eqt{E05} as
\begin{equation}
\left| A \left( \nu_{\ell} \to \nu_{\ell'} \right) \right|
=
\left|
\sum_{i=2,3} U_{\ell' i} U_{\ell i}^*
\left[
\exp\left( \displaystyle -i { \Delta m^2_{i1} R \over 2 p } \right)
- 1 \right]
+ \delta_{\ell'\ell}
- \Omega_{\ell'\ell}
\right|
\;,
\label{E06}
\end{equation}
where
$ \Delta m^2_{ij} \equiv m_i^2 - m_j^2 $,
$R\simeq t$ is the distance
between the source and the detector,
\begin{equation}
\Omega_{\ell'\ell}
=
\sum_{a\ge4} U_{\ell' a} U_{\ell a}^*
\;.
\label{E07}
\end{equation}
and the relation \eqtm{E03} was used.
If all the squared mass differences
are so small that
\hbox{$ \displaystyle { \Delta m^2_{i1} R \over p } \ll 1 $},
then
the transition amplitude is constant
and given by
\begin{equation}
\left| A \left( \nu_{\ell} \to \nu_{\ell'} \right) \right|
\simeq
\left|
\delta_{\ell'\ell} - \Omega_{\ell'\ell}
\right|
=
\left| \braket{ \nu_{\ell'} }{ \nu_{\ell} } \right|
\;.
\label{E08}
\end{equation}
So the part of the transition amplitude
which does not depend on $R$ and $p$
is connected with the non-orthogonality
of the states in \eqt{E04},
which describe neutrinos taking part
in weak interactions
in the case of mixing
among light and heavy Majorana particles.
Different phenomenological
aspects of this non-orthogonality
have been considered in
Refs.[\ref{Glashow87},\ref{Langacker88}].
Here we discuss the implications
of neutrino masses, mixing and non-orthogonality
for neutrino oscillations
\footnote{
Let us notice that in
the case of non-orthogonality
the usual notion of flavour neutrinos
loose its meaning.
For example,
a neutrino which is created with a muon
and is described by the state $\ket{\nu_{\mu}}$
can produce besides a muon also an electron or a tau.
}.

The transition amplitude
in \eqt{E06}
contains many unknown parameters.
In the following
we make some general assumptions
about the neutrino masses
and the elements of the mixing matrix
which reduce the number of relevant parameters.
We assume that there is a hierarchy
of light neutrino masses
\begin{equation}
m_1 \ll m_2 \ll m_3
\label{E09}
\;.
\end{equation}
The masses of all known fundamental fermions
satisfy this type of hierarchy.
If the neutrino masses are generated
through the see-saw mechanism,
then
$ \displaystyle m_{i} \simeq { m_{f_{i}}^2 \over M_{i} } $,
where,
for each generation $i$,
$ m_{f_{i}} $
is the mass of the up-quark or charged lepton
and
$ M_{i} $
is the mass of the heavy Majorana fermion.
In this case,
the hierarchy
in \eqt{E09}
is obviously satisfied.
We assume also
that the masses
$ m_1 $ and $ m_2 $
are too small
to be relevant for terrestrial neutrino oscillation experiments.
The masses
$ m_1 $ and $ m_2 $
could be responsible for the MSW resonant transition
of solar neutrinos,
which can explain all existing experimental data,
including the new GALLEX data [\ref{SolarData}].

There are some indications at present that the mass $ m_3 $
could be in the electronvolt region.
One indication come from
the analysis of the recent COBE data
and the observations
of the large scale distribution of galaxies [\ref{COBE}].
Another indication
comes from the MSW explanation
of the solar neutrino data [\ref{SolarData}]
together with a see-saw formula for the
neutrino masses.

Taking into account all these arguments,
the probability of
$ \stackrel{(-)}{\nu}_{\mu} \to \stackrel{(-)}{\nu}_{\tau} $
transition is given by
\begin{equation}
\begin{array}{lll} \displaystyle \hei
P \left( \stackrel{(-)}{\nu}_{\mu} \to \stackrel{(-)}{\nu}_{\tau} \right)
& = & \displaystyle
2
\left| U_{\tau3} \right|^2
\left| U_{\mu 3} \right|^2
\left[
1 - \cos\left( { \Delta m^2_{31} R \over 2 p } \right)
\right]
\\ \displaystyle \hei
& & \displaystyle
- 2
\left| U_{\tau3} \right|
\left| U_{\mu 3} \right|
\left| \Omega_{\tau\mu} \right|
\left[
\cos\left( { \Delta m^2_{31} R \over 2 p } \mp \chi \right)
-
\cos \chi
\right]
\\ \displaystyle
& & \displaystyle \hei
+
\left| \Omega_{\tau\mu} \right|^2
\;,
\end{array}
\label{E10}
\end{equation}
where
\begin{equation}
\chi
\equiv
{\rm A r g}
\left\{ U_{\tau3} U_{\mu3}^* \Omega_{\tau\mu}^* \right\}
\label{E11}
\;.
\end{equation}

The following comments are in order:

\begin{enumerate}

\item
If there is a hierarchy of couplings
in the lepton sector
\begin{equation}
\left| U_{e 3} \right|^2
\ll
\left| U_{\mu 3} \right|^2
\ll
\left| U_{\tau3} \right|^2
\label{E12}
\;,
\end{equation}
analogous to the hierarchy in the quark sector,
then the
$ \nu_{\mu} \to \nu_{\tau} $
transition probability
is the largest one
[\ref{Bilenky92}].

\item
It can be seen from \eqt{E10}
that
if there is no $ \Omega $ term in the amplitude,
the
$ \nu_{\mu} \to \nu_{\tau} $
and
$ \bar\nu_{\mu} \to \bar\nu_{\tau} $
transition probabilities are equal.
If there is a mixing among light and heavy Majorana particles
and CP is violated in the lepton sector
(the mixing matrix is complex),
the
$ \nu_{\mu} \to \nu_{\tau} $
and
$ \bar\nu_{\mu} \to \bar\nu_{\tau} $
transition probabilities could be different.
So a comparison between
these transition probabilities
could be a test
for the mixing of light neutrinos with heavy Majorana particles
and CP violation in the lepton sector.

\item
Besides the usual
$ \displaystyle { R \over p } $-dependent term
$ \displaystyle
\left[ 1 - \cos\left( { \Delta m^2_{31} R \over 2 p } \right) \right]
$,
the transition probability in \eqt{E10}
contains an additional
$ \displaystyle { R \over p } $-dependent term
with a phase shift $\chi$.
Notice that
this additional term is proportional to
$ \left| U_{\mu3} \right| $,
which is much bigger than the
$ \left| U_{\mu3} \right|^2 $
in the usual term
if there is a hierarchy relation \eqt{E12}.
So a measurement of the energy dependence
of the transition probability
allows in principle
to get information about the mixing
of the light neutrinos
with heavy Majorana particles.

If the squared mass difference
$ \Delta m^2_{31} $
is so small that
$ \displaystyle { \Delta m^2_{31} R \over p } \ll 1 $,
then
\begin{equation}
P \left( \stackrel{(-)}{\nu}_{\mu} \to \stackrel{(-)}{\nu}_{\tau} \right)
\simeq
\left| \Omega_{\tau\mu} \right|^2
=
\left| \braket{ \nu_{\tau} }{ \nu_{\mu} } \right|^2
\;.
\label{E14}
\end{equation}
In this case the transition probability
does not depend on
$ \displaystyle { R \over p } $
and it is determined by the non-orthogonality
of the $ \nu_{\mu} $ and $ \nu_{\tau} $ states.
Such a situation was discussed in
Ref.[\ref{Langacker88}].

\item
An upper bound for the value of
$ \left| \Omega_{\tau\mu} \right| $
in \eqt{E10}
is given by the Schwartz inequality
\begin{equation}
\left| \Omega_{\tau\mu} \right| \le
\sqrt{
\left| \Omega_{\mu\mu} \right|
\left| \Omega_{\tau\tau} \right|
}
\;.
\label{E15}
\end{equation}
Information about the values of
$ \left| \Omega_{\mu\mu} \right| $
and
$ \left| \Omega_{\tau\tau} \right| $
can be obtained from the analysis
of different weak interaction processes.
Such analysis was done in
Refs.[\ref{Langacker88},\ref{Bilenky90}].
Using the values obtained in
Ref.[\ref{Bilenky90}]
we have the following $2\sigma$
upper bounds
\begin{equation}
\begin{array}{l} \displaystyle \hei
\left| \Omega_{\mu\mu} \right| \alt 4 \times 10^{-3}
\;,
\\ \displaystyle \hei
\left| \Omega_{\tau\tau} \right| \alt 2 \times 10^{-1}
\;.
\end{array}
\label{E16}
\end{equation}
    From \eqts{E15} and \eqtm{E16}
we have
\begin{equation}
\left| \Omega_{\tau\mu} \right| \alt 2.8 \times 10^{-2}
\;.
\label{E17}
\end{equation}

The upper bound of the value of
$ \left| \Omega_{\tau\mu} \right| $
can be estimated also from the LEP measurement
of the number of neutrino flavours.
In the case under consideration
the number of neutrino flavours
is given by
[\ref{Bilenky90}]
\begin{equation}
N_{\nu}
=
{ \Gamma_{\rm inv} \over \Gamma_{0} }
=
3 - 2 \sum_{\ell} \Omega_{\ell\ell}
+ \sum_{\ell\ell'} \left| \Omega_{\ell\ell'} \right|^2
\;.
\label{E18}
\end{equation}
Using the relation
\begin{equation}
\left| \Omega_{\tau\mu} \right|
\le
{1\over\sqrt{2}}
\sum_{\ell} \Omega_{\ell\ell}
\;.
\label{E19}
\end{equation}
and the latest LEP data
$ N_\nu = 2.99 \pm 0.04 $ [\ref{LEP92}],
we obtain
\begin{equation}
\left| \Omega_{\tau\mu} \right| \alt 3 \times 10^{-2}
\;.
\label{E20}
\end{equation}

Let us compare
the value of the coefficients
of the first and second terms
of the transition probability in \eqt{E10}.
Using \eqts{E17} and \eqtm{E20} for their ratio we have
\begin{equation}
R =
{
\left| \Omega_{\tau\mu} \right|
\over
\left| U_{\tau3} \right|
\left| U_{\mu3} \right|
}
\le
{
3 \times 10^{-2}
\over
\left| U_{\tau3} \right|
\left| U_{\mu3} \right|
}
\;.
\label{E21}
\end{equation}

In many models (see for example Ref.[\ref{Langacker91}])
the elements of the lepton mixing matrix $U$
are approximately equal to
the elements of the Cabibbo-Kobayashi-Maskawa
mixing matrix of quarks.
In this case
$ \left| U_{\tau3} \right| \left| U_{\mu3} \right|
\simeq (4.3 \pm 0.7) \times 10^{-2} $
[\ref{PDG92}],
leading to
\begin{equation}
R \alt 1
\;.
\label{E22}
\end{equation}

In a recent paper [\ref{Ellis92}]
the elements of the lepton mixing matrix
were calculated
in the context of a flipped SU(5) model.
The value of
$ \left| U_{\tau3} \right| \left| U_{\mu3} \right| $
in this model
depends on the top quark mass $m_t$
and is predicted to lie in the interval
\begin{equation}
0.9 \times 10^{-2} \ (m_t=150\GeV)
\le
\left| U_{\tau3} \right| \left| U_{\mu3} \right|
\le
2 \times 10^{-2} \ (m_t=90\GeV)
\;.
\label{E23}
\end{equation}
    From \eqt{E23}
the upper bound for the ratio $R$ is given by
\begin{equation}
\begin{array}{lcl} \displaystyle \hei
R \le 3.3
& \hskip0.5truecm & \displaystyle
{\rm for} \qquad
m_t=150\GeV
\;,
\\ \displaystyle \hei
R \le 1.5
& \hskip0.5truecm & \displaystyle
{\rm for} \qquad
m_t=90\GeV
\;.
\end{array}
\label{E24}
\end{equation}

These estimates show
that the coefficient of the second term
of the transition probability in \eqt{E10}
could be comparable
to the coefficient of the ``main'' term.

    From our estimation also follows
that the third term of the transition probability in \eqt{E10},
which does not depend on
$ \displaystyle { R \over p } $,
is bounded by
\begin{equation}
\left| \Omega_{\tau\mu} \right|^2 \alt 10^{-3}
\;.
\label{E25}
\end{equation}
Therefore,
if the neutrino squared mass difference
is so small that the first and second terms in \eqt{E10}
vanish,
the third term,
which is due to the non-orthogonality
of the neutrino states,
could give
a constant $ \nu_\mu \to \nu_\tau $ transition.

\end{enumerate}

In conclusion,
our considerations show
that a detailed investigation
of
$ \nu_\mu \to \nu_\tau $
and
$ \bar\nu_\mu \to \bar\nu_\tau $
transitions
in future CERN and Fermilab experiments
could allow to get informations
not only about neutrino masses and mixing
but also about the possibility
of a mixing of the light neutrinos
with heavy Majorana particles.

\vskip1cm

It is a pleasure for us
to thank prof. V. de Alfaro
for fruitful discussions.
\newpage
{\Large\bf References }

\begin{list}{[\therefs]}{\usecounter{refs}}

\item\label{SolarData}
R. Davis, in {\it Neutrino '88},
Proceedings of the $XIII^{\rm th}$
International Conference on Neutrino
Physics and Astrophysics,
ed. J. Scheps et al. (World Scientific 1989) p. 518;
K.S.~Hirata et al.,
{\it Phys. Rev. Lett.} {\bf 65} (1990) 1297,
{\it Phys. Rev. Lett.} {\bf 65} (1990) 1301;
GALLEX Collaboration,
{\it Phys. Lett.} B {\bf 285} (1992) 390.

\item\label{AtmData}
K.S. Hirata et al.,
{\it Phys. Lett.} B {\bf 280} (1992) 146;
D. Casper et al.,
{\it Phys. Rev. Lett.} {\bf 66}, 2561 (1991).

\item\label{NewSolarExp}
For a review see:
D. Sinclair,
{\it Nucl. Phys.} B (Proc. Suppl.) {\bf 19} (1991) 100.

\item\label{Schneps92}
J. Schneps,
Talk presented at the
$15^{\rm th}$ International Conference on Neutrino Physics
and Astrophysics (NEUTRINO 92),
Granada, Spain, June 1992.

\item\label{CHORUS}
CHORUS Collaboration,
CERN-SPSC/90-42 (1990).

\item\label{NOMAD}
NOMAD Collaboration,
CERN-SPSC/91-21 (1991).

\item\label{FNAL}
R. Bernstein et al.,
Neutrino Physics after the Main Injector Upgrade,
FNAL 1991.

\item\label{Bilenky78}
See the reviews:
S.M. Bilenky and B. Pontecorvo,
{\it Phys. Rep.} {\bf 41} (1978) 225;
S.M. Bilenky and S.T. Petcov,
{\it Rev. Mod. Phys.} {\bf 59} (1987) 671.

\item\label{LEP92}
The LEP Collaborations,
{\it Phys. Lett.} B {\bf 276} (1992) 247.

\item\label{Langacker91}
P. Langacker,
Lectures presented at TASI-90,
Boulder, June 1990,
UPR 0470T.

\item\label{GRSY}
M. Gell-Mann, P. Ramond and R. Slansky,
in {\it Supergravity},
ed. F. van Nieuwenhuizen and D. Freedman,
North Holland, Amsterdam 1979, p. 315;
T. Yanagita,
{\it Prog. Theor. Phys.} B {\bf 135} (1978) 66.

\item\label{Glashow87}
J.~Bernabeu et al.,
{\it Phys. Lett.} B {\bf 187} (1987) 303;
S.L. Glashow
{\it Phys. Lett.} B {\bf 187} (1987) 367;
J.W.F. Valle,
{\it Phys. Lett.} B {\bf 199} (1987) 432.

\item\label{Langacker88}
P. Langacker and D. London,
{\it Phys. Rev.} D {\bf 38} (1988) 886,
{\it Phys. Rev.} D {\bf 38} (1988) 907.

\item\label{COBE}
M. Davis, F.J. Summers and D. Schlegel,
{\it Nature} {\bf 359} (1992) 393;
A.N. Taylor and M. Rowan-Robinson,
{\it Nature} {\bf 359} (1992) 396.

\item\label{Bilenky92}
S.M. Bilenky, M. Fabbrichesi and S.T. Petcov,
{\it Phys. Lett.} B {\bf 276} (1992) 223.

\item\label{Bilenky90}
S.M. Bilenky, W. Grimus and H. Neufield,
{\it Phys. Lett.} B {\bf 252} (1990) 119.

\item\label{PDG92}
Review of Particle Properties,
{\it Phys. Rev.} D {\bf 45} (1992) Part II.

\item\label{Ellis92}
J. Ellis, J.L. Lopez and V. Nanopoulos,
CERN-TH.6569/92, June 1992.

\end{list}
\end{document}